\documentclass[usenatbib]{mnras}
\usepackage[dvips]{color}
\usepackage{ulem}
\usepackage{longtable}
\usepackage{url}

\usepackage{graphicx}
\usepackage{lscape}

\def\arcmin{\hbox{$^\prime$}}
\def\arcsec{\hbox{$^{\prime\prime}$}}

\def\flux{erg s$^{-1}$ cm$^{-2}$}
\def\lum{erg s$^{-1}$}

\def\aap{A\&A}

\def\apjs{ApJS}

\def\apjl{ApJLett}

\def\igr{IGR\,J18027-2016}

\def\a{$^{\mbox{\small a}}$}
\def\b{$^{\mbox{\small b}}$}
\def\c{$^{\mbox{\small c}}$}

\title[CRSF in \igr\ with {\it NuSTAR}]
{{\it NuSTAR} observations of the supergiant X-ray pulsar \igr: accretion from the
stellar wind and possible cyclotron absorption line}
\author[Lutovinov et al.]{Alexander\,A.\,Lutovinov$^{1,2}$\thanks{E-mail: aal@iki.rssi.ru},
Sergey\,S.\,Tsygankov$^{3}$, Konstantin\,A.\,Postnov$^{4}$,
\newauthor Roman\,A.\,Krivonos$^{1}$, Sergey\,V.\,Molkov$^{1}$ and John\,A.\,Tomsick$^{5}$
\\
$^{1}$ Space Research Institute of the Russian Academy of Sciences, Profsoyuznaya
str. 84/32, Moscow, 117997, Russia\\
$^2$ Moscow Institute of Physics and Technology, Institutskiy per. 9, Dolgoprudny,
Moscow Region, 141700, Russia \\
$^3$ Tuorla Observatory, Department of Physics and Astronomy, University of Turku,
  V\"ais\"al\"antie 20, FI-21500 Piikki\"o, Finland \\
$^4$ Moscow M.V. Lomonosov State University, Sternberg Astronomical Institute,
119992 Moscow, Russia\\
$^5$ Space Sciences Laboratory, 7 Gauss Way, University of California, Berkeley,
CA 94720-7450, USA
}

\begin{document}

\date{Accepted .... Received ...}

\pagerange{\pageref{firstpage}--\pageref{lastpage}} \pubyear{2016}

\maketitle

\label{firstpage}

\begin{abstract}
{We report on the first focused hard X-ray view of the absorbed supergiant
system \igr\ performed with the {\it NuSTAR} observatory. The pulsations are
clearly detected with a period of $P_{\rm spin}=139.866(1)$ s and a pulse
fraction of about 50-60\% at energies from 3 to 80 keV. The source
demonstrates an approximately constant X-ray luminosity on a time scale of
more than dozen years with an average spin-down rate of $\dot
P\simeq6\times10^{-10}$ s s$^{-1}$. This behaviour of the pulsar can be
explained in terms of the wind accretion model in the settling regime. The
detailed spectral analysis at energies above 10 keV was performed for the
first time and revealed a possible cyclotron absorption feature at energy
$\sim 23$~keV. This energy corresponds to the magnetic field
$B\simeq3\times10^{12}$ G at the surface of the neutron star, which is
typical for X-ray pulsars.}
\end{abstract}


\section{Introduction}

\igr\ was discovered by the {\it INTEGRAL} observatory during deep
observations of the Galactic Center region as a faint persistent hard X-ray
source \citep{2004AstL...30..382R}. Follow-up observations performed with the
{\it XMM-Newton} observatory in soft X-rays allowed for an improvement in the
determination of the source position, enabling near-infrared observations with
the {\it NTT}/ESO telescope and establishing the nature of its companion as a
supergiant star of B1Ib type \citep{2010A&A...510A..61T} at a distance of 12.4 kpc.
\citet{2011A&A...532A.124M} classified the optical star as B0-B1 I, which is
broadly in agreement with the above mentioned type.

Using archival data from the {\it BeppoSAX} observatory, \citet{2003ApJ...596L..63A}
revealed that the serendipitous source SAX\,J1802.7-2017 is spatially associated
with \igr. These authors showed that this source is an X-ray pulsar with a spin
period of 139.6 s, residing in a binary system with an orbital period of 4.6 days.
Later, the orbital parameters of the system were improved by \citet{2005A&A...439..255H}
and \citet{2015A&A...577A.130F}.

The average spectrum of \igr, measured in a wide energy band
\citep{2005A&A...430..997L}, demonstrates a cutoff at high energies, which is
typical for X-ray pulsars \citep[see, e.g.,][]{2005AstL...31..729F}. In
addition, a relatively high absorption value, $N_{\rm H}=6.8\times10^{22}$
cm$^{-2}$, was revealed at low energies \citep{2005A&A...439..255H}, and
the source was classified as an absorbed binary system with a
supergiant companion \citep{2015A&ARv..23....2W}.  No search for a
cyclotron absorption line and no pulse phase resolved spectroscopic
study for \igr\ has been performed to date.

In this paper we report on results of observations of \igr\ collected with the
{\it NuSTAR} observatory and {\it Swift}/XRT telescope in Aug 2015. The main
purpose of these observations was to characterize the broadband spectrum
with high accuracy and to search for a cyclotron absorption line.

\section{Observations and data reduction}

The {\it Nuclear Spectroscopic Telescope Array} ({\it NuSTAR}) has two identical
X-ray telescope modules, each equipped with independent mirror systems and
focal plane detector units, also referred to as FPMA and FPMB
\citep{2013ApJ...770..103H}. The unique multilayered coating of the
grazing-incidence Wolter-I optics provides X-ray imaging in the energy
range $3-79$~keV with an angular resolution of 18\arcsec\ (FWHM)
and 58\arcsec\ (HPD). Each telescope has a field-of-view (FOV) of
$\sim13'\times13'$. CdZnTe pixel detectors of FPMA and FPMB provide spectral
resolution of 400 eV (FWHM) at 10 keV.

\begin{figure}
\includegraphics[width=1.0\columnwidth,bb=50 225 560 569,clip]{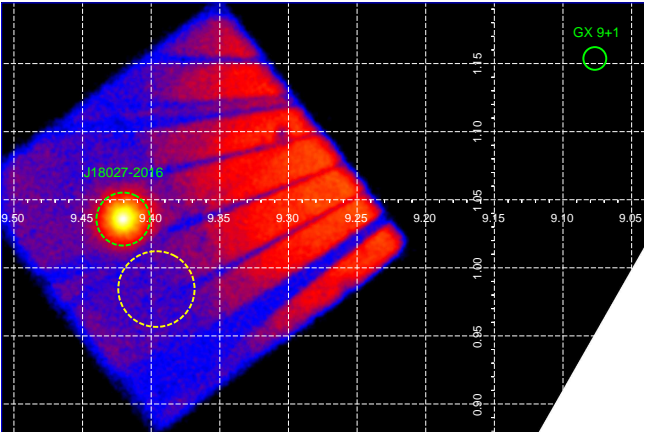}
\caption{The {\it NuSTAR} image of \igr\ in the $3-79$~keV energy band obtained
with FPMA module (the picture for another module FPMB is practically
identical). Green and yellow dashed circles denote source and background
extraction regions, respectively. The ghost-ray (GR) contamination is revealed
by a pattern of characteristic radial streaks in the image. The GR photons
originate from GX\,9+1, whose position is marked by the solid green circle.
The white grid indicates the Galactic coordinate system in
degrees.}\label{nuimage}
\end{figure}

\igr\ was observed with {\it NuSTAR} on August 27, 2015 from MJD\,57,261.0141
to 57,261.9922 (ObsID 30101049002, the total data span length is about 78 ks)
during Cycle~1 of the Guest Observer Program with a total on-target exposure
of $\sim36$~ks ($\sim32.6$~ks after the dead-time correction and GX\,9+1
flare removal, see below). The presence of bright nearby sources can
contaminate the {\it NuSTAR} observations due to the aperture stray light --
unfocused X-rays reaching the {\it NuSTAR} detectors. The contamination that
we see is often called ghost-rays (GR), which are photons that are reflected
from only upper or lower X-ray mirror sections (so-called single scattering
photons, see, e.g., \citealt{2014ApJ...792...48W, 2014ApJ...781..107K,
2015ApJ...814...94M}). The bright low-mass X-ray binary system GX\,9+1 is
located in close proximity to \igr\ (at an angular distance of 21.5\arcmin).
Therefore, initially (at the observation planning stage), we checked for
possible contamination of GR photons from this source to the useful signal.
According to {\it NuSTAR's} experience with GRs from the very bright system
4U\,1630-47 \citep{2014ApJ...791...68B}, the direct GRs are observed within a
$\sim20$\arcmin\ radius circle, which made observations of \igr\ possible.
Indeed, Fig.\,\ref{nuimage} demonstrates that there is no a strong GR
contamination from the persistent emission of GX\,9+1 at the position of
\igr. However, a strong X-ray flare was detected in the light curve of \igr.
Timing and spectral characteristics of this event are typical for Type I
X-ray bursts observed from some accreting neutron stars. The burst is also
clearly present in a light curve extracted from the background data, which
are fully dominated by the GX\,9+1 GR contamination. Therefore, we suggest
that the burst originated from GX\,9+1, and we removed the corresponding part
of the data from the following analysis. Note that the exposure accumulated
during the burst (300 sec in total) is too short to identify a characteristic
pattern of ghost-rays.

\begin{figure}
\includegraphics[width=0.9\columnwidth,bb=51 272 547 672]{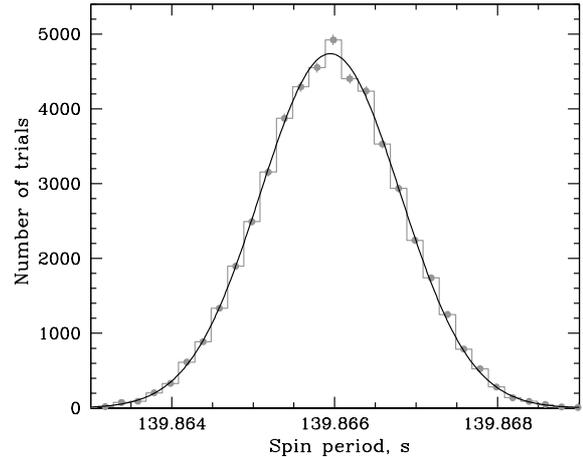}
\caption{Distribution of periods obtained for $5\times10^4$ trial light
curves (histogram) and its best-fit approximation with the Gaussian (solid
line).}\label{fig:efsearch}
\end{figure}

We processed the raw observational data to produce cleaned event files for
the FPMA and FPMB modules using the standard {\it NuSTAR} Data Analysis
Software ({\sc NuSTARDAS}) v1.6.0 provided under {\sc HEASOFT v6.19} with
CALDB version 20160502. Using the {\sc nuproducts} routine, we extracted
source counts for spectral and timing analysis within 70\arcsec\ of the
centroid position (dashed green circle in Fig.\,\ref{nuimage}), which
corresponds to $\sim80\%$ of the enclosed PSF fraction
\citep{2015ApJS..220....8M}. Taking into account possible GR contamination,
we extracted the background spectrum and light curves in the region with a
radius of 100\arcsec (dashed yellow circle), located at nearly same distance
from GX\,9+1 (its position is indicated by the solid green circle in
Fig.\,\ref{nuimage}). Finally, after background subtraction the total
number of registered photons from the source is about $\simeq1.36\times10^5$
for both modules.

\begin{figure}
\includegraphics[width=0.95\columnwidth,bb=70 140 500 710,clip]{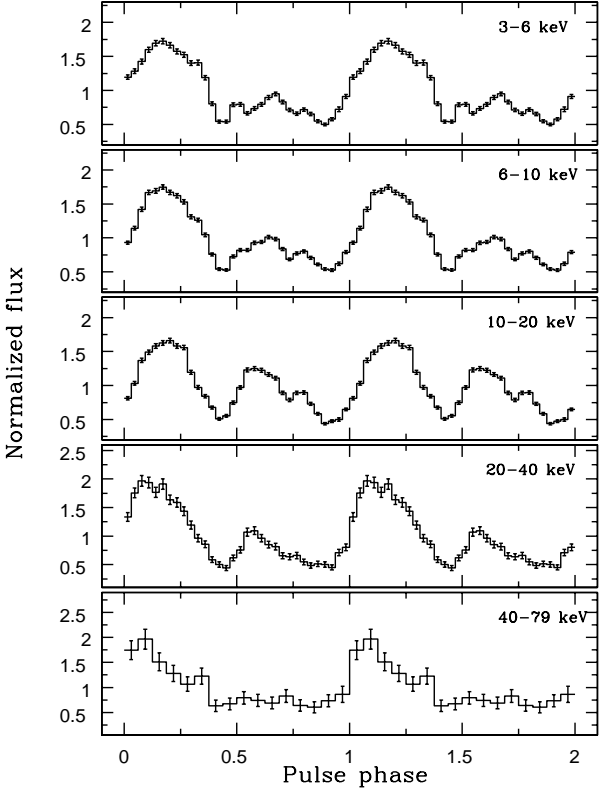}
\caption{Pulse profiles of \igr\ in five energy bands 3--6, 6--10,
  10--20, 20--40, and 40--79 keV (from top to bottom).  The profiles
  are normalized by the mean flux in each energy band and plotted twice for
  clarity.}\label{fig:pprof}
\end{figure}

For timing analysis and pulse phase-resolved spectroscopy, the data were
corrected for the barycenter of the Solar System with the {\sc barycorr}
command and for the binary system motion using orbital parameters from
\citet{2015A&A...577A.130F}.

The {\it Swift}/XRT instrument observed \igr\ simultaneously with the {\it
NuSTAR} observatory on August 27, 2015 (MJD\,$57261.6525-57261.7159$, ObsID
00081660001) with a total exposure of $\sim2$ ks. The {\it Swift}/XRT data
were used to supplement the {\it NuSTAR} data in the soft energy band and to
better determine the level of absorption. These data were reduced
using appropriate software\footnote{http://swift.gsfc.nasa.gov}.

All of the spectra were then grouped to have more than 20 counts per
bin using the {\it grppha} tool from the {\sc HEAsoft} package. The final
data analysis (timing and spectral) was performed with the {\sc FTOOLS 6.7}
software package.

\section{Timing analysis}

The {\it NuSTAR} observatory provides data with very good time and energy
resolution, allowing us to investigate temporal properties of \igr\ at
energies above 10 keV in detail for the first time. In general, the source
light curve measured by {\it NuSTAR} demonstrates two types of
variability on time scales longer than the pulse period and shorter than
the orbital period: smooth changes of the source intensity during the
observation and relatively quick variations (up to $25-30$\%) on the time
scale of several thousand seconds. It can be naturally expected that both
types of variations are connected with gradual changes of the stellar wind
density over the orbit of the neutron star and its local inhomogeneities.

\begin{figure}
\includegraphics[width=0.95\columnwidth,bb=45 310 515 730]{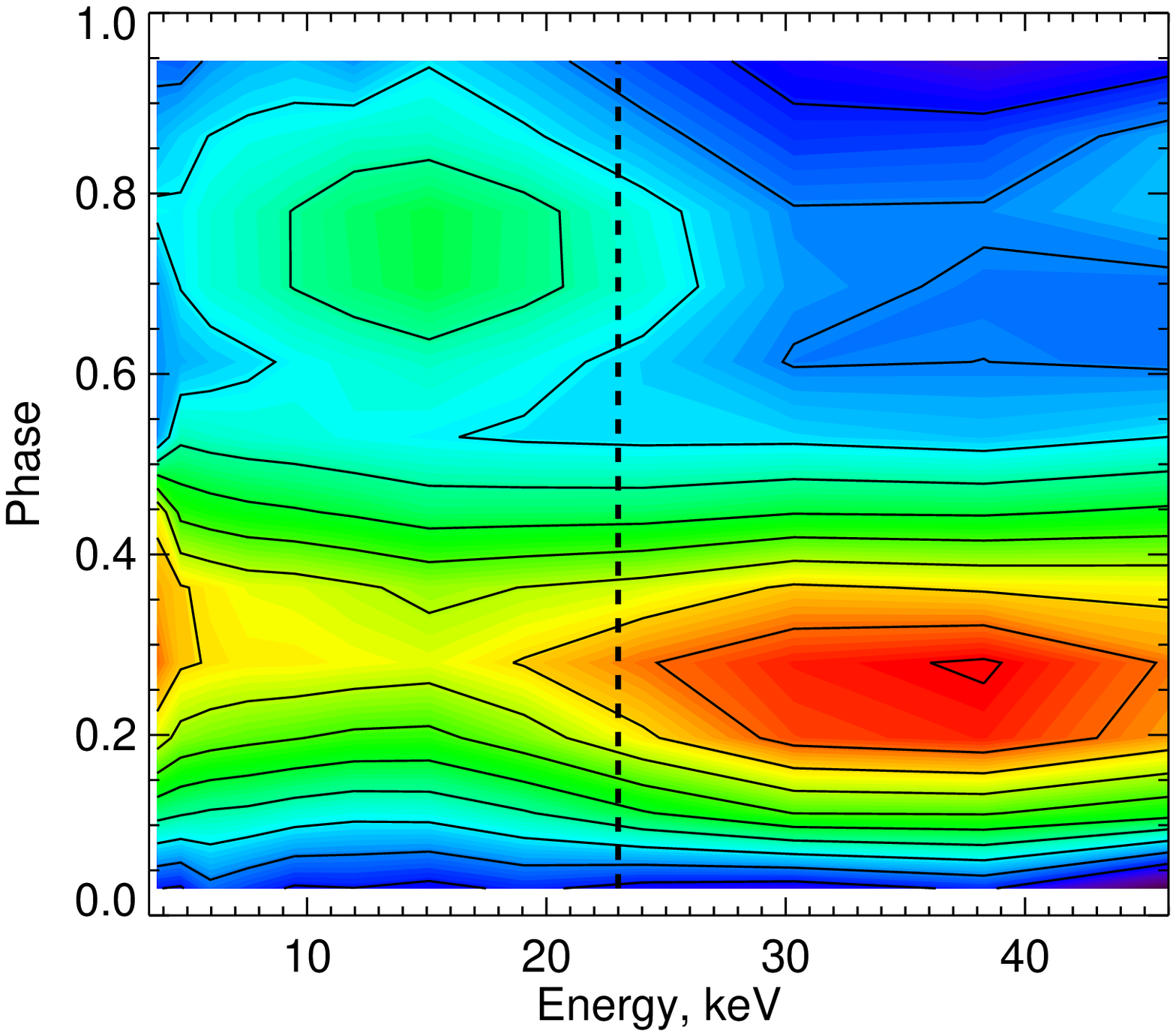}
\caption{``Energy -- pulse phase'' distribution of intensity for \igr. The
dashed vertical line shows the position of the cyclotron line revealed in the
source spectrum. The profiles were also normalized by the mean flux in each
energy band. Solid contours represent levels of the equal normalized flux
from 0.5 to 1.85 with steps of 0.15.}\label{fig:2dpprof}
\end{figure}

The pulse period and its uncertainty were determined using the combined light
curve from both modules FPMA and FPMB \citep[see][for the description of the
procedure]{2015ApJ...809..140K} in the wide energy band (3--79 keV) with 1 s
time bins and corrected both for the motion of Earth around Sun and the
neutron star in the binary system. In general, a correct determination of the
pulse period, signal amplitude, their uncertainties using the epoch folding
technique is not trivial task and different approaches have been proposed
\citep[see, e.g.][]{1987A&A...180..275L, 1996A&AS..117..197L,
2011A&A...529A..30D}. Here, to estimate an uncertainty on the pulse period we
generated $5\times10^4$ trial light curves (using the statistics from the
original one, provided with the {\sc nuproducts} package) and determined the
pulse period value in each of them using the epoch folding technique. We used
the {\sc efsearch} procedure from the {\sc FTOOLS} package with 24 trial
profile phase bins. The distribution of periods obtained has a smooth shape,
which can be well describe with a Gaussian profile (see
Fig.\,\ref{fig:efsearch}). The mean and standard deviation corresponds
approximately to the proper pulse period and its $1\sigma$ uncertainty,
respectively \citep[see, e.g.,][for details]{2013AstL...39..375B}. The final
value of the pulse period obtained by this analysis is $P_{\rm
spin}=139.866\pm0.001$ s, where the uncertainty corresponds to $1\sigma$
confidence level.

The obtained value of $P_{\rm spin}$ is larger by about $\simeq0.25$ s
compared with the previous measurements made with the {\it BeppoSAX} and {\it
XMM-Newton} observatories \citep{2005A&A...439..255H}, resulting in the mean
spin-down rate of $\dot P\simeq6\times10^{-10}$ s s$^{-1}$.

\begin{figure}
\includegraphics[width=0.98\columnwidth,bb=20 270 515 675]{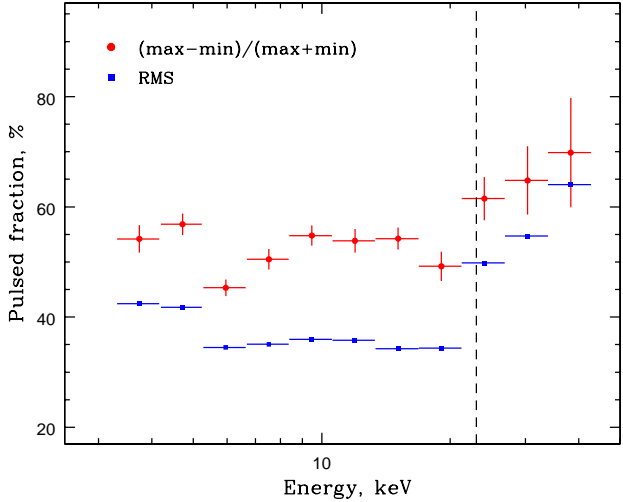}
\caption{Dependence of the pulsed fraction of \igr\ on energy calculated in
two different ways (see text for details). The dashed line shows the position
of the cyclotron line that may be present in the source spectrum.}\label{fig:ppfr}
\end{figure}

The pulse profiles in five different energy bands 3--6, 6--10, 10--20, 20--40,
and 40--79 keV (from top to bottom) are shown in Fig.\,\ref{fig:pprof}. In all
energy bands, it has a double peak shape with different relative intensities
of the peaks.

The evolution of the pulse profile with energy can be illustrated by the
``energy -- pulse phase'' distribution of the intensity made possible by {\it
NuSTAR}’s energy resolution (see Fig.\,\ref{fig:2dpprof}). These pulse
profiles were also normalized by the mean flux in each energy band. Details
of the technique of the construction of such a 2D-distribution are described
by \citet{2009AstL...35..433L} and have already been successfully used in a
number of works. From Fig.\,\ref{fig:2dpprof}, it is clear that the second
peak has its highest relative intensity between $\sim10$ and $\sim20$ keV. At
higher energies, the first peak completely dominates the profile. It is
important to note that these two energy ranges are divided by the possible
cyclotron absorption line revealed in the source spectrum (see
Section~\ref{sec:spec}). Its position is shown with the dashed line in
Fig.\,\ref{fig:2dpprof}.

The pulsed fraction below the cyclotron line energy depends on energy only
slightly with a value of 50--60\% (see Fig. \ref{fig:ppfr}). At higher
energies one can see the gradual increase of the pulsed fraction, typical for
the majority of X-ray pulsars \citep{2009AstL...35..433L}.

To avoid possible biases due to the pulse profile peculiarities or
statistics, we used two different definitions of the pulsed fraction. The
``standard'' one is
$\mathrm{PF}=(F_\mathrm{max}-F_\mathrm{min})/(F_\mathrm{max}+F_\mathrm{min})$,
where $F_\mathrm{max}$ and $F_\mathrm{min}$ are maximum and minimum fluxes in
the pulse profile, respectively. Red circles in Fig.\,\ref{fig:ppfr} show the
pulsed fraction defined by this manner. Another way to characterize the pulse
profile intensity variations is to use the relative Root Mean Square (RMS),
which can be calculated by the following equation:

\begin{equation}\label{rms}
RMS=\frac{\Big(\frac{1}{N}\sum_{i=1}^N(P_i-\langle P\rangle)^2\Big)^{\frac{1}{2}}}{\langle P\rangle},
\end{equation}
where $P_i$ is the background-corrected count rate in a given bin of
the pulse profile, $\langle P\rangle$ is the profile-averaged count
rate, and $N$ is the total number of phase bins in the profile.

The overall behaviour of the pulsed fraction calculated using the different
approaches agrees very well with the only difference being a slightly lower
absolute value for the RMS (blue squares in Fig. \ref{fig:ppfr}). This
dependence shows two distinct features: the increase of the pulsed fraction
with energy and its peculiar change at $22-24$ keV. Note that there is another
similar change of the pulsed fraction around 6-7 keV probably connected with
the fluorescent iron emission line.

\begin{figure}
\includegraphics[width=0.95\columnwidth,bb=46 152 550 690,clip]{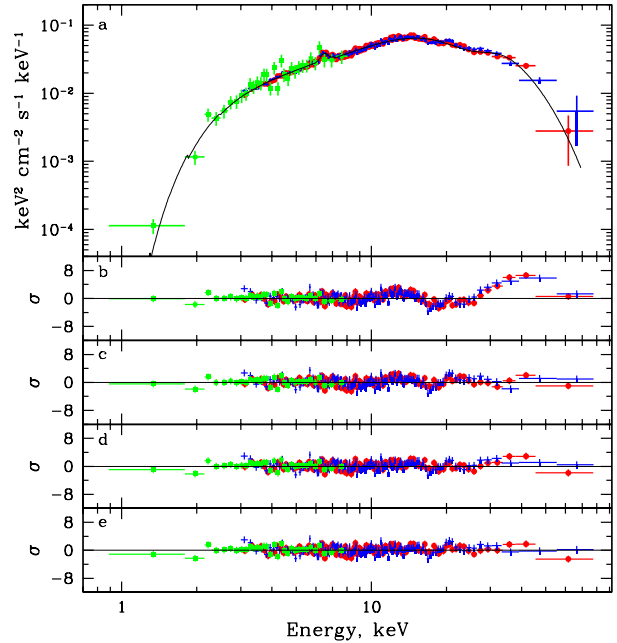}
\caption{(a) Pulse phase averaged spectrum of \igr\ obtained with the {\it
NuSTAR} observatory and the {\it Swift}/XRT instrument. Red circles and blue
crosses correspond to data from FPMA and FPMB, respectively; green points --
to the XRT data. The black line shows the best fit model. (b-e) Residuals
from different models (see text for details). The spectrum and residuals
were rebinned in the figure to make the deviations more readily apparent.
}\label{fig:avspec}
\end{figure}

\section{Spectral analysis}
\label{sec:spec}

\begin{table*}
\caption{Best-fit parameters of the \igr\ spectrum with different models.}\label{tablspec}
\begin{tabular}{lllll}
\hline
\hline
Parameter & Comptt  & Comptt+Gabs & PowHigh\a  &  PowHigh+Gabs \\

\hline
$C1$                               &$1.014\pm0.006$ &$1.013\pm0.006$ &$1.013\pm0.006$ & $1.013\pm0.006$ \\
$C2$                               & $0.87\pm0.04$   & $0.86\pm0.04$ &$0.86\pm0.04$ & $0.86\pm0.04$ \\
$N_{\rm H}$, $\times10^{22}$ cm$^{-2}$ &$3.55\pm0.30$& $3.22\pm0.30$ &$3.36\pm0.22$ & $3.23\pm0.21$ \\
$kT_{\rm 0}$, keV                  & $0.61\pm0.09 $  & $0.70\pm0.09$ &-- & -- \\
$kT_{\rm }$, keV                   & $5.91\pm0.04 $  & $6.86\pm0.08$ &-- & -- \\
$\tau$                             & $7.59\pm0.13 $  & $7.41\pm0.14$ & -- & -- \\
Photon index                       & --              & --            & $0.89\pm0.02$  & $0.87\pm0.02$ \\
$E_{\rm cut}$, keV                 & --              & --            & $12.49\pm0.13$ & $12.93\pm0.25$ \\
$E_{\rm fold}$, keV                & --              & --            & $11.10\pm0.18$ & $11.56\pm0.31$ \\
$\tau_{\rm cycl}$                  & --              &$0.53^{+0.08}_{-0.07}$ & --     & $0.16^{+0.07}_{-0.04}$\\
$E_{\rm cycl}$, keV                & --              & $24.26\pm0.49$& --             & $23.2\pm1.0$  \\
$\sigma_{\rm cycl}$, keV           & --              &$5.93^{+0.46}_{-0.42}$& --      & $5.35^{+1.13}_{-0.75}$\\
$E_{\rm Fe}$, keV                  & $6.46\pm0.03$   & $6.45\pm0.03$ & $6.46\pm0.03$  & $6.46\pm0.03$  \\
$\sigma_{\rm Fe}$, keV             & $0.2$ (fix)     & $0.2$ (fix)   & $0.2$ (fix)    & $0.2$ (fix)  \\
$EW$, eV                           &$106^{+23}_{-11}$&$121^{+29}_{-9}$& $117^{+13}_{-13}$&$118^{+12}_{-8}$ \\[2mm]
Flux (3--79 keV)\b                 & \multicolumn{4}{c}{$(1.695^{+0.010}_{-0.016})\times10^{-10}$ \flux} \\
Luminosity (3--79 keV)\c           & \multicolumn{4}{c}{$(3.107^{+0.019}_{-0.029})\times10^{36}$ \lum} \\
$\chi^2$ (d.o.f)                   & $1572.3 (1127)$ & $1229.1 (1124)$ &  $1211.4 (1127)$ & $1172.5 (1124)$ \\

\hline
\end{tabular}
\vspace{3mm}

\begin{tabular}{ll}
\a & the {\sc powerlaw*highcut} in the {\sc XSPEC} package \\
\b & measured flux \\
\c & for a distance of 12.4 kpc  (Torrej{\'o}n et al., 2010) \\
\end{tabular}
\end{table*}

The working energy range of the {\it NuSTAR} observatory (3--79 keV), and its
unprecedented sensitivity at $>10$ keV makes the observatory an ideal
instrument to search for cyclotron resonant scattering features (or, in other
words, cyclotron absorption lines) in the spectra of X-ray pulsars \citep[for
more recent examples see, e.g.,][]{2016ApJ...823..146B,tsygankov2016}.
Moreover, the quality of the {\it NuSTAR} data is so high that the simple
theoretical models usually used for the modelling of spectra of X-ray pulsars
are often not able to approximate them satisfactorily, and we were met with
this problem in the analysis of the \igr\ spectrum.

As mentioned above, the spectrum of \igr\ is typical for accreting X-ray
pulsars.  In particular, it demonstrates an exponential cutoff at high
energies (Fig.\,\ref{fig:avspec}a). Therefore, as a first step, we attempted
to approximate it with several commonly used models: a power law with an
exponential cutoff ({\sc cutoffpl} in the {\sc XSPEC} package), the NPEX
model, the broken power law model \citep[see, e.g.,][]{2005A&A...439..255H},
thermal inverse Comptonization ({\sc comptt} in the {\sc XSPEC} package), a
power law with a high energy cutoff ({\sc powerlaw*highcut} in the {\sc
XSPEC} package, \citealt{1983ApJ...270..711W}). To take into account the
absorption in the system, the {\sc phabs} model was used with the abundances
from \citet{1989GeCoA..53..197A}.

The source and background spectra from both the FPMA and FPMB modules of
{\it NuSTAR} as well as from the {\it Swift}/XRT telescope were used for
simultaneous fitting. To take into account the uncertainty in the instrument
calibrations as well as the lack of full simultaneity of observations by
{\it NuSTAR} and {\it Swift} (the former observed \igr\ for nearly the
whole day, while the latter observed for only 2 ks), cross-calibration
constants between them were included in all spectral models (the
$C1$ and $C2$ constants from Table\,\ref{tablspec} correspond to the
cross-calibrations of the FPMB module and the XRT telescope to
the FPMA module, respectively).

None of the models that we used describe the spectrum very well -- typically,
there are strong deviations at low energies and a depression around 20 keV
with a complex shape. Moreover, two models (the broken power law and {\sc
powerlaw*highcut}) suffer from the abrupt breaks at the break and cutoff
energies, respectively. This can lead to artificial narrow line features
around these energies \citep[see, e.g.,][]{shtykovsky16}. Therefore, in the
following analysis, we used the {\sc comptt} model, which we consider to be
the most reliable one.

Residuals between the source spectrum and the {\sc comptt} model are shown in
Fig.\,\ref{fig:avspec}b. In comparison with other models, the {\sc comptt}
model approximates the soft part of the spectrum quite well and has a clear
physical meaning. At the same time, Fig.\,\ref{fig:avspec}b also shows a
peculiarity in the spectrum -- a deficit of photons around 20 keV, leading
to a high $\chi^2$ value (Table\,\ref{tablspec}).

To investigate this feature, we added an absorption component to the model in
the form of the {\sc gabs} model in the {\sc XSPEC} package.  It led to a
significant improvement of the fit quality and $\chi^2$ value
(Table\,\ref{tablspec}). The corresponding residuals are shown in
Fig.\,\ref{fig:avspec}c, and the line centroid energy is $E_{cycl}\simeq24.3$
keV. We interpret this absorption feature as a possible cyclotron absorption
line. Two models, {\sc gabs} and {\sc cyclabs} in the {\sc XSPEC} package,
are usually used to approximate absorption lines.  While both models provide
an adequate description of the data, the cyclotron line energy derived from
the {\sc gabs} model is systematically higher by several keV than the energy
derived from the {\sc cyclabs} model \citep[see,
e.g.,][]{2012MNRAS.421.2407T,2015MNRAS.448.2175L}. For \igr\ and the {\sc
cyclabs} model, the cyclotron line energy would be $E_{cycl}\simeq20.2$ keV.
Note that some wavelike structure is still present in residuals, which can be
attributed to physical causes or to inperfections in the model.

Besides the cyclotron absorption line, a strong emission feature associated
with the fluorescent iron emission line is clearly visible near 6.4 keV. To
take it into account, a corresponding component in the Gaussian form was
added to the model. The final best fit parameters as well as the source
intensity and luminosity are summarized in Table\,\ref{tablspec}.

\begin{figure}
\includegraphics[width=0.9\columnwidth,bb=55 250 460 680]{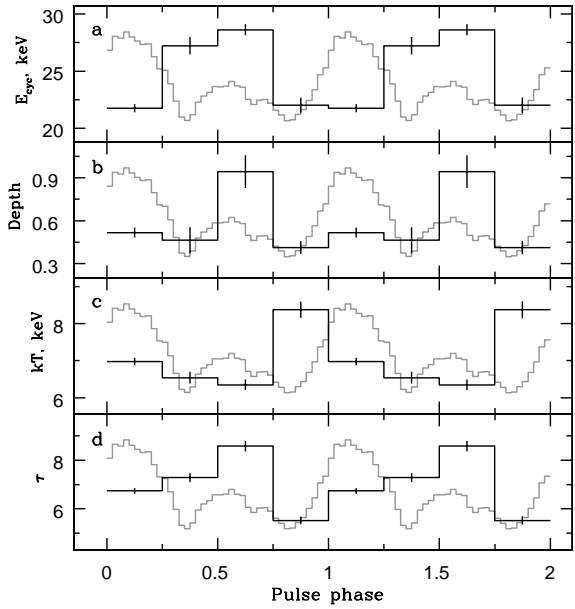}
\caption{Spectral parameters of the comptonization model as a function
of pulse phase. The black histogram in the panels represents: (a) cyclotron
line energy, (b) cyclotron line depth, (c) temperature and (d) optical depth
in the {\sc comptt} model. The grey line in each panel shows the pulse
profile in a wide energy range.}\label{fig:parpuls}
\end{figure}

In order to trace the evolution of the source spectrum and its parameters
over the pulse period, we carried out pulse phase-resolved spectroscopy.
Taking into account the relative faintness of the source (its flux is only
about 4-5 mCrab), the spin period was divided into four wide phase bins,
roughly corresponding to the main and secondary peaks (Fig.\,\ref{fig:parpuls}).
Such a division allows us to obtain spectra with good statistical quality
at each phase and to determine the spectral parameters.

To approximate pulse phase-resolved spectra, we used the same model as for
the averaged spectrum.  However, the temperature of the seed photons $kT_{\rm
0}$ and the cyclotron line width were fixed to the values measured for the
averaged spectrum because the temperature and width are poorly constrained
due to the limitations of the spectral quality of the spectra.  The column
density was also fixed because only {\it NuSTAR} data (i.e., $>3$ keV) were
used for the phase-resolved spectroscopy.

Variations of the spectral parameters over the pulse are shown in
Fig.\,\ref{fig:parpuls}. For better visualization, the pulse profile of \igr\
in a wide energy band 3-79 keV is plotted in each panel by the grey
histogram. The continuum parameters ($kT$ and $\tau$) change slightly with
the pulse phase, demonstrating a possible anticorrelation between them. In
contrast, the cyclotron line energy demonstrates strong dependence of the
pulse phase -- in the main peak, it is around $E_{\rm cycl}\simeq22$ keV,
while in the secondary peak, it increases to $E_{\rm cycl}\simeq28$ keV.
Considering the relatively small width of the line ($\sim5.5$ keV), it can
lead to the wave-like structure in the residuals (see
Fig.\,\ref{fig:avspec}c).

Finally, it is necessary to note that, formally, the best approximation of
the \igr\ spectrum is given when the {\sc powerlaw*highecut} model is used for
the continuum. This model was proposed by
\citet{1983ApJ...270..711W} to approximate spectra of accreting X-ray
pulsars. Despite the artificiality and presence of the abrupt break at the
cutoff energy, it describes spectra obtained with the majority of X-ray
observatories more or less adequately. The significant increase of the
data quality in hard X-rays with the {\it NuSTAR} observatory raises a
question about the physically based models for X-ray pulsar emission as,
e.g., use of the {\sc powerlaw*highcut} model can lead to artificial
narrow absorption lines near the cutoff energy. Nevertheless, the \igr\
spectrum is an example where this model works and describes it without
leaving additional artificial features (see
Table\,\ref{tablspec}). The $\chi^2$ value for this model is significantly
lower than for {\sc comptt} and is comparable with {\sc comptt + gabs}.
However, even in this case, there is still some noticeable deficit of
photons around 20 keV, which can be approximated by the adding of the {\sc
gabs} component to the model (see Fig.\,\ref{fig:avspec}d,e). The improvement
of the fit quality for this case is not so significant as for the {\sc
comptt} model. Therefore, we performed extensive Monte-Carlo simulations
($10^5$ trials) to estimate the significance of the line detection and
found that it is about 4$\sigma$ for the {\sc powerlaw*highecut} continuum
model.

\begin{figure}
\includegraphics[width=0.9\columnwidth,bb=55 150 460 680]{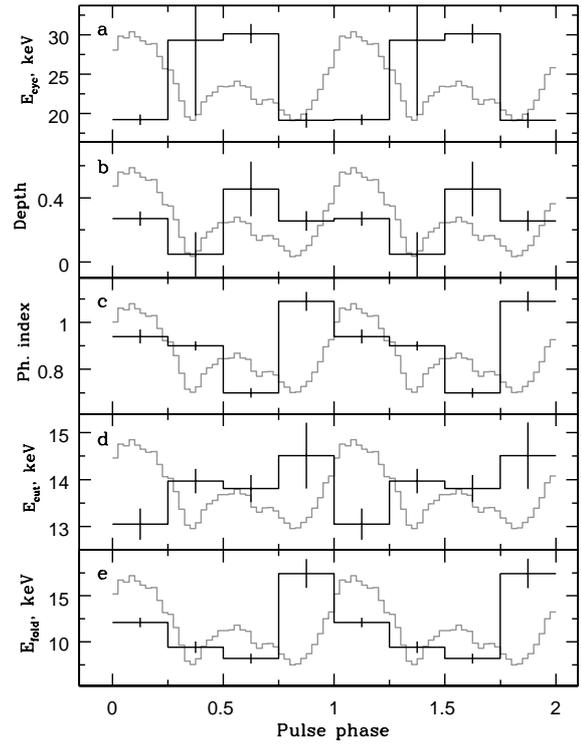}
\caption{The same as in Fig.\,\ref{fig:parpuls}, but for the {\sc
powerlaw*highcut} model of the continuum. The black histogram in the panels represents: (a)
cyclotron line energy, (b) cyclotron line depth, (c) photon index, (d) cutoff
energy and (e) folding energy.}\label{fig:parpuls2}
\end{figure}

Results of the pulse phase-resolved spectroscopy for the {\sc
powerlaw*highecut} continuum model are shown in Fig.\,\ref{fig:parpuls2} in
the same manner as in Fig.\,\ref{fig:parpuls}. The cyclotron line is not
significantly detected in all phase bins, but its energy as a function of
pulse phase is qualitatively similar to the behavior seen when using the
comptonization continuum model -- in the main peak, it is around
$E_{\rm cycl}\simeq20$ keV, while in the secondary peak, it increases up to
$E_{\rm cycl}\simeq30$ keV. Again, the cyclotron line width and column
density were fixed to values from the pulse averaged spectrum.

\section{Discussion and conclusion}

In this paper, we report results of the {\it NuSTAR} observations of the
absorbed supergiant system \igr. They can be summarized as follows: 1) the system
demonstrates approximately constant X-ray luminosity on a time scale of more
than a dozen years; 2) during this time interval, the pulsar spun down with a
rate of $\dot P\simeq6\times10^{-10}$ s s$^{-1}$; 3) the possible
presence of a cyclotron absorption line at $\sim 23$~keV is found in
the source X-ray spectrum.

The observed increase of the rotation period of the neutron star in \igr\ can
be explained in terms of the wind accretion model in a settling regime
\citep{2012MNRAS.420..216S,2014EPJWC..6402001S}, which can be realized for
slowly rotating magnetized neutron stars at X-ray luminosities below $\sim
4\times 10^{36}$~erg s$^{-1}$. The observed stable X-ray luminosity suggests
that the X-ray pulsar has reached its equilibrium period, which, in
this model, depends on the binary orbital period $P_b$, the stellar wind velocity $v_w$,
the neutron star magnetic moment $\mu$ and the mass accretion rate $\dot M$:

\begin{equation}
\label{e:Peq}
P_{eq}\simeq 940[\mathrm{s}]\mu_{30}^{12/11}\left(\frac{P_b}{10\mathrm{d}}\right)
\dot M_{16}^{-4/11}v_8^4\,,
\end{equation}

\noindent where the characteristic parameters are $\mu_{30}\equiv
\mu/10^{30}[\mathrm{G\,cm}^3]$, $\dot M_{16}\equiv \dot
M/10^{16}[\mathrm{g\,s}^{-1}]$, $v_8\equiv v_w/10^8[\mathrm{cm\,s}^{-1}]$. A
very strong dependence on the wind velocity suggests that it is more reliable
to estimate the wind velocity by inverting this formula.  From the
observed X-ray luminosity (see Table 1) we find $\dot M_{16}\simeq 3$, and
assuming that the absorption feature is the cyclotron line, we find the
neutron star dipole magnetic moment $\mu_{30}\simeq 2$. Therefore, assuming
that the observed pulsar period is close to its equilibrium value,
$P_{eq}=P_{spin}=139.6$~s, we can estimate the wind velocity $v_8\sim 0.7$. We
stress that this estimate depends only very weakly on parameters ($\dot M$ and
$\mu$ and other model-dependent numerical coefficients). Note that this
rather low value is very close to the wind velocity measured in the
prototypical persistent HMXB with OB supergiant Vela X-1
\citep{2016A&A...591A..26G}, suggesting similarity between the two sources.

In the model of settling quasi-spherical accretion, the neutron star close to
equilibrium can exhibit either spin-up or spin-down, depending on whether the
actual $\dot M$ is above or below the critical mass accretion rate, $\dot
M_{eq}$ (see Eq. (2) in \citealt{2014EPJWC..6402002P}). For the parameters of \igr,
we find $\dot M_{eq}\simeq 3.6\times 10^{16}$~g s$^{-1}$, i.e. indeed
spin-down of the neutron star is possible in this system. The observed value
of the negative torque acting on the neutron star in \igr\, $\dot\omega\simeq
2\times 10^{-13}$~rad~s$^{-2}$, does not exceed the maximum possible negative
torque, $\omega_{sd,max}\simeq 10^{-12}$~rad~s$^{-2}$, for the parameters of \igr\
(see Eq. (6) in \citealt{2014EPJWC..6402002P}). Thus, the observed steady-state
spin-down is consistent with expectations from the settling accretion model.

As discussed in Section~\ref{sec:spec}, the spectral analysis revealed
the possible presence of a cyclotron absorption line in the spectrum of \igr\ at
energies of $\sim23-24$ keV. An additional independent test for the
existence of this feature comes from the timing properties of the source.
Namely, the pulse profile and pulsed fraction dependencies on the energy have
prominent features near the same energy. Particularly, the relative intensity
of two peaks in the profile changes around 22-24 keV (see
Fig.\,\ref{fig:2dpprof}). Such behaviour was shown to be typical for
another well known X-ray pulsar V\,0332+53 with a well established cyclotron
feature \citep{2006MNRAS.371...19T}. The observed non-monotonic dependence of
the pulsed fraction on energy (see Fig.\,\ref{fig:ppfr}) is also typical for
pulsars with cyclotron lines \citep{2009AstL...35..433L,2009A&A...498..825F}.
These additional observational facts indirectly support the presence of the
cyclotron absorption line in the \igr\ spectrum despite of its low
significance for the {\sc powerlaw*highcut} continuum model.

The measured X-ray luminosity in \igr, $L_x\simeq 3\times 10^{36}$ \lum,
suggests that accretion onto the neutron star occurs in the subcritical
regime where the radiation plays a secondary role \citep[see,
e.g.][]{1976MNRAS.175..395B,2015MNRAS.447.1847M}.  In this case, the accretion
flow is decelerated in a collisionless shock formed at some height above the
neutron star surface \citep{1982ApJ...257..733L,2004AstL...30..309B}. The
formation of a cyclotron line downstream of the shock occurs in the resonance
layer in the inhomogeneous magnetic field, so the line profile can be more
complicated than the simple Doppler-broadened line
\citep{1996ASSL..204.....Z}. For example, depending on the geometry, the line
may have a flat bottom or show emission wings. The complex shape of the
residuals shown in Fig.\,\ref{fig:avspec}, which are obtained assuming a
Gaussian form of the line, may suggest a complicated line profile or may be
evidence for the presence of complicated magnetic field structure near
the surface of the neutron star, as discussed, for example,
in \cite{2012MNRAS.420..720M}. Moreover, as the optical thickness of the
flow at the resonant energies is very high even at low mass accretion rate,
the cyclotron line can be also formed in the accretion channel above the
shock region \citep{2015MNRAS.454.2714M}. In this case, one would expect
the cyclotron absorption feature to be redshifted relative to the actual
cyclotron energy. This interesting result should be studied in the future
with deeper observations.

\section*{Acknowledgments}

This work was supported by the Russian Science Foundation (grant
14-22-00271). Work of PK is partially supported by RSF grant 14-12-00657.
Authors thanks to the anonymous referee for useful comments.
The research has made use of data obtained with {\it NuSTAR}, a project led
by Caltech, funded by NASA and managed by NASA/JPL, and has utilized the
NUSTARDAS software package, jointly developed by the ASDC (Italy) and Caltech
(USA).

\bsp    
\label{lastpage}
\end{document}